\documentclass[
    ,final            
    ,sort&compress    
    ,numberedheadings 
  ]
  {aipproc}

\usepackage{hyperref}           
\usepackage[latin1]{inputenc}   

\layoutstyle{8x11single}

\providecommand{\email}[1]{E-mail:
\href{mailto:#1}{\textnormal{\texttt{#1}}}}

\begin{document}

\title{Research on candidates for non-cosmological redshifts}


\author{M. L\'opez-Corredoira}{
  address={Instituto de Astrof\'\i sica de Canarias (IAC),
  E-38200 La Laguna (S/C de Tenerife, SPAIN), \email{martinlc@iac.es}}
}

\author{C. M. Guti\'errez}{
  address={IAC, \email{cgc@iac.es}}
}

\begin{abstract}
The paradox of apparent optical associations of galaxies with very different 
redshifts, the so-called anomalous redshift problem, is around 35 years old, 
but is still without a clear solution and is surprisingly ignored by most of 
the astronomical community. Statistical correlations among the positions of 
these galaxies have been pointed out by several authors. 
Gravitational lensing by dark matter has been proposed as the 
cause of these correlations, although this seems to be insufficient to explain 
them and does not work at all for correlations with the brightest and 
nearest galaxies. Some of these cases may be just fortuitous associations in 
which background objects are close in the sky to a foreground galaxy, although 
the statistical mean correlations remain to be explained and some lone objects 
have very small probabilities of being a projection of background objects. 

The sample of discordant redshift associations given in Arp's atlas is indeed 
quite large, and most of the objects remain to be analysed thoroughly. For about
5 years, we have been running a project to observe some of these cases in 
detail, and some new anomalies have been added to those already known; For 
instance, in some exotic configurations such as NGC 7603 or NEQ3, which can even 
show bridges connecting four object with very different redshifts,
and the probability for this to be a projection of background sources 
is very low. Not only QSOs but also emission-line galaxies in general are found to take part 
in this kind of event.
Other cases are analyzed: MCG 7-25-46, GC 0248+430,
B2 1637+29, VV172 and Stephan's Quintet and, in some of them, it is probable
that the associations stem from a background 
projection, although some other low probability features 
are present in some of these systems which remain to be understood.
\end{abstract}

\maketitle


\section{The problem and the observations which give rise to it}

The problem of apparent optical associations of galaxies with 
very different redshifts, the so-called anomalous 
redshifts\cite{Arp87,Nar89,Arp03}, is old but still alive. 
Although surprisingly ignored by most of the astronomical 
community, over the last 35-40
years the amount of evidence in favour of it has been steadily increasing,
both in the study of individual cases and 
statistically\cite{Chu84,Bur96,Bur01}: an excess of high redshift sources near
low redshift galaxies, positive and very significant cross
correlations between surveys of galaxies and QSOs, an excess of
pairs of QSOs with very different redshifts, etc.
These observations and analysis suggest that there may be some objects whose
redshift is not due to the cosmological expansion supposed by
the Big Bang model, but that the redshift is intrinsic to the object (anomalous).
This motivated us to begin our own independent observations and analyses.
In this paper we will pay attention to the study of a few examples
of anomalous redshift candidates, particularly objects we
have recently studied.

\subsection{Phenomenology}

Among the facts suggesting that there is something anomalous with
some systems of objects, something which suggests that they are associated
rather than being a fortuitous projection of objects with different distances,
are:

\begin{itemize}

\item There is an excess of galaxies or QSOs with high redshifts near
the centre of nearby galaxies. In some cases, the QSOs are only a few arcseconds
away from the centre of the galaxies. This is observed in the aforementioned 
cross-correlations of surveys but also in some individual cases.
Examples considered here are NEQ3, which has a QSO/HII galaxy pair  with redshift 0.19 
and another HII galaxy only 2.6$''$ from it with $z=0.22$, the
VV 172 chain and Stephan's Quintet with a closer discordant object.
Other cases in the literature include NGC 613, NGC 3079, NGC 3842, 
NGC 7319 (in Stephan's Quintet, with a QSO 8$''$ distant from the centre),
2237+0305 (with a QSO less than 0.3$''$ from the centre), etc.

\item Filaments/bridges/arms apparently connecting objects with different
redshifts. This is the case, for instance, of NGC 7603, NEQ3, MCG 7-25-46 
which will be described below. In these cases, the origin of such 
filaments should be related to the interaction of 
different galaxies, and only galaxies with 
different redshift appear to be appropriate candidates for such interactions. 
Moreover, the probability of chance projections of background/foreground objects 
on to the filament, as observed in some cases, is very low. 
Other cases in the literature include  AM 2052-221, NGC 4319+Mrk 205, Mrk273, 
NGC 3067+3C232 (in radio), NGC 622, NGC 1232, NGC 4151, etc.

\item Distance indicators which suggest that some galaxies or QSOs are much
closer than indicated by their redshifts. An example discussed in this paper is NGC 7603: 
the spectra of two HII galaxies with high redshifts is typical of dwarf HII
galaxies, whose luminosity points to possibly  much lower distances than indicated
by their redshift. Other cases in the literature are
the anomalously large sizes of NGC 262 or NGC 309 if their redshifts 
are taken as distance indicators.

\item The alignment of sources with different redshifts, 
which suggest that they have a common origin, and that the direction of alignment
is the direction of ejection, as proposed by theories claiming
that the redshifts are intrinsic rather than cosmological. This happens
with some configurations of QSOs. Examples are shown here with GC 0248+430,
B2 1637+29. Other cases in the literature include: 1130+106, 3C212, NGC 4258,
NGC 2639, NGC 4235, NGC 5985, etc.

\item No absorption lines in QSOs that are supposed to be located behind foreground galaxies.
We have not observed any of these targets, but there are several in the
literature, e.g. PKS 0454+036.

\item Morphological evidence of interaction such as 
star-forming regions, distorted shapes, asymmetries in rotation curves, etc.
We have paid no attention to this kind of evidence---perhaps some distortions
in NEQ3---but there are several
cases in the literature, e.g. NGC 450+UGC 807.

\end{itemize}

In the following sections, we illustrate some of those cases 
presenting apparent anomalies observed by us over the last 5 years. This
is a small sample among the vast literature on the topic that has accumulated
during the last
35-40 years\cite{Arp87,Nar89,Bur96,Arp03}, especially by H. C. Arp, 
M. Burbidge, G. Burbidge, J. Sulentic and others. 
A surprising fact regarding our observations is that we have observed only
about a dozen systems\footnote{The reason is mainly because we obtained
only a few nights of observing time on 2--4 m telescopes.
Most of the observations were done
in 2001--2. In subsequent applications, no time was obtained 
in spite of our having published several papers in major astronomical journals 
on the topic and demonstrated the utility
of our observations within this field of anomalous redshifts.}, 
that were known to present some 
apparent anomaly, and nearly in half of them we have found some new apparent 
anomaly. Were  we just lucky, or is it that these old and new apparent anomalies 
reflect the existence of genuinely non-cosmological redshifts?

\section{NGC 7603}

\begin{figure}
\includegraphics[width=6cm]{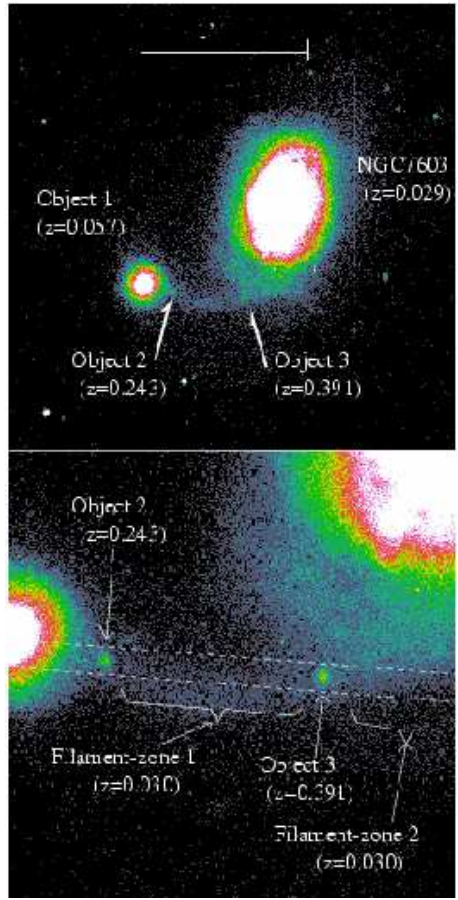} 
\caption{NGC 7603 and the surrounding
field. $R$-filter, taken on the 2.5 m Nordic Optical Telescope (La Palma, Spain).
Reproduction of Fig. 1 of \cite{Lop02}.}
\label{Fig:f1}
\end{figure}

Fig. \ref{Fig:f1} shows an example of a system suggesting an anomalous
redshift. The main galaxy, NGC 7603, is a broad line Seyfert I galaxy 
with $z=0.0295$ and $B$ = 14.04 mag \cite{Vau91}.
This galaxy has been studied mainly  in relation to its 
variability\cite{Kol00}. The Balmer lines are unusually broad, 
show a very complex structure and are blueshifted relative to the 
local `rest frame' of the AGN by between 1000 and 2000 km/s. Less 
than 5\% of AGN show such characteristics. Such lines are more common in 
radio-loud quasars, where one sees ejected synchrotron lobes. It also shows 
unusually strong Fe II emission for an AGN with such broad 
lines \cite{Goo89,Kol00}.

A fact that has attracted attention 
is the proximity of NGC 7603B, a spiral galaxy with higher 
redshift ($z=0.0569$) located 59 arcseconds to the SE of 
NGC 7603 \cite{Arp71,Arp75,Sha86}. 
The angular proximity of both galaxies and the apparently luminous
connection between them makes the system an important example of 
a possible anomalous redshift association. 
Arp \cite{Arp71,Arp75,Arp80} has claimed that the compact member, NGC 7603B, 
was somehow ejected from the bigger object.

Moreover, there
are also two objects overimposed on the filament
 apparently connecting both galaxies. 
We identified several emission lines
in the spectra of the two knots, and from the emission lines of H$\alpha$, 
H$\beta$, OII, doublet OIII, NII-6584 \AA \ 
we determined their redshifts to be 0.394$\pm 0.002$ and 
0.245$\pm 0.002$ for the objects closest to and farthest from NGC 7603 
respectively\cite{Lop02,Lop04}.
B-magnitudes corrected for extinction (due to the filament) are
respectively 21.1$\pm 1.1$  and 22.1$\pm 1.1$ \cite{Lop04}.
According to the line ratios, these objects are HII-galaxies but are quite peculiar:
the very intense H$_\alpha $
[equivalent width: EW(H$_\alpha )\approx $80 \AA \ and 160\AA \ resp.] 
is indicative of a vigorous star-formation galaxy. 
Only $\sim 2$\% and $\sim 1$ \% resp. of the normal HII-galaxies have a so 
high EW(H$_\alpha$) \cite{Car01}.
However, if they were dwarf HII-galaxies, 
these high EWs would be within the normally expected values.
If we consider the redshifts as indicators of 
distance, the respective absolute magnitudes would be\cite{Lop04}:
$M_V=-21.5\pm 0.8$ and $-18.9\pm 0.8$. However, if we consider 
an anomalous intrinsic redshift case 
(in such a case, in order to derive the distance, we set $z=0.03$), 
the results are: $M_V=-15.2\pm 0.8$ and $-13.9\pm 0.8$ resp.\cite{Lop04}. 
In this second case, they would be on the faint tail of the HII-galaxies, 
type II \cite{Tel95}; they would be dwarf galaxies, 
``tidal dwarfs'', and this would explain the observed strong 
star formation ratio: objects with low luminosity have higher 
EW(H$_\alpha $)\cite{Car01}. Of course, this would imply that we have
non-cosmological redshifts.

\begin{figure}
\includegraphics[width=9cm,angle=-90]{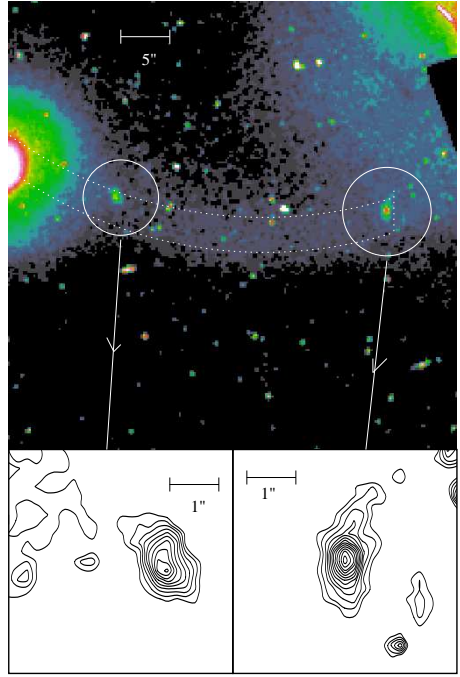} 
\caption{HST image in  F606W of the region centred on the 
filament between NGC 7603 and NGC 7603B. Also shown are the contours of 
the two objects in the filament. Note that there are many bad pixels/cosmic 
rays in the images that do not correspond to any object. The PSF is 
$\sim 0.1$ arcsec. Dotted lines show the area (around 140 arcsec$^2$) 
that we consider ``filament'' for the calculation of 
probabilities. Reproduction of Fig. 6 of \cite{Lop04}.}
\label{hst}
\end{figure}

Figure \ref{hst} shows the field from the Hubble Space Telescope\cite{Lop04}. 
The field is centred on the filament
between NGC 7603 and NGC 7603B and clearly  shows  the two objects within it.
The FWHM of both objects is between 0.3 and 0.4 arcseconds, 
which is very small to be measured with a ground telescope and 1 arcsecond seeing, 
 and seems to indicate that they are extended rather than point-like 
objects. The two HII-galaxies in the filament are apparently a little 
deformed, although the significance is not too high (the two lowest 
isocontours in Fig. \ref{hst} are $\sim 2\sigma $ and $\sim 3.5 \sigma$ 
respectively above the average flux in the region). 
The tail of the object close to NGC 7603 in the northern part is warped 
 towards NGC 7603, and the other object 
has a faint apparent tail in the northern part but this tail 
is less significant. It could indicate that the material in the 
filament interacts with the galaxies.

\begin{figure}
\includegraphics[width=8cm,angle=0]{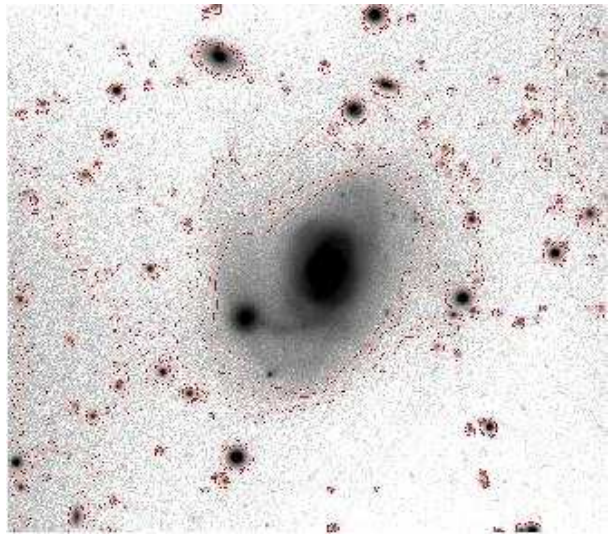} 
\caption{A gray scale and contour image in the R band of the region around the 
galaxy NGC 7603. The contours correspond to isophotes 24.8, 25.3 and 26.2
mag/arcsec$^2$. Taken on the 2.5 m NOT  (La Palma, Spain).
Reproduction of Fig. 1 of \cite{Lop04}.}
\label{Rhalo}
\end{figure}

The  NGC 7603-NGC 7603B system appears to be surrounded by a diffuse halo that 
we have been able to delineate down 
to 26.2 mag/arcsec$^2$ in the r-band Sloan filter (Fig. \ref{Rhalo}). 
Although this halo seems to be associated mostly with NGC 7603, 
it is not symmetric with respect to this galaxy. There is evidence of a
fainter extension tail to the north. The last isophote of the halo is 
also asymmetric to the west, possibly including a counter arm
of the bright filament between NGC 7603 and NGC 7603B.
The halo+filament between NGC 7603 and NGC 7603B shows up clearly and has a
maximum brightness of 22.9 mag/arcsec$^2$ in the Sloan r-band,
while the halo near the filament has a brightness of
23.4 mag/arcsec$^2$. Therefore, the filament alone is approximately 
24.0 mag/arcsec$^2$. Another diffuse structure is seen also apparently connecting 
 NGC 7603 and NGC 7603B, and situated to the south of
the main filament. A point-like object 
situated to the southeast  of this tail is a local star\cite{Lop04}. 

From several absorption lines we estimated the redshift of the filament 
apparently connecting NGC 7603 and NGC 7603B as $z=0.030$ \cite{Lop02}, very
similar to the redshift of NGC 7603 and probably associated with this galaxy. 
There is an absence of H$_\alpha $ emission lines in NGC 7603B and in the 
filament\cite{Sha86,Lop04}.
The non-detection of emission lines is not proof against the existence of a 
physical connection. In interactions and ejections with a larger galaxy, 
the gas is often stripped away from a stellar system\cite{Ros01}, 
so the lack of emission lines could be taken as
an indication of interaction rather than non-interaction 
(against what is pointed out by \cite{Sha86}). Fig. \ref{Rhalo} shows that
NGC 7603 and its filament are apparently distorted by significant tidal 
interaction. The  existence of the filament itself is also a possible sign 
of tidal interaction or debris from satellite disruption\cite{Joh01}. 
The fainter southern filament and the red fringe embedded in NGC 7603 
reinforces the   tidal debris scenario. But where is the companion that
interacts with NGC 7603?
There is a galaxy with similar redshift, one magnitude fainter, 
and 10.3 arcminutes from NGC 7603: NGC 7589; or B231533.01-000313.1, 
three magnitudes fainter and 12.6 arcminutes of distance. However, all of 
them are in the opposite direction to the filament (in the west instead of 
the east). There are two closer galaxies 2--3 arcminutes to the north
(SDSS J231901.1+001651.8 and SDSS J231855.52+001619)
which are 4--5 magnitudes fainter but they both have a redshift of $z=0.077$
(measured by the Sloan Digital Sky Survey).
We do not find any appropriate candidate for the interaction in the field
down to the limiting magnitude for spectroscopy in SDSS survey: either they
are very far and in the opposite direction or they have very different
redshifts. Maybe this lost companion is very faint, but then it would
have too low a mass or too high a distance to produce a   
24 mag/arcsec$^2$ filament.
Johnston et al.\cite{Joh01} in their eq. (11)/fig. 6 calculate the expected 
surface brightness magnitude in such cases. Assuming $t>1$ Gyr 
a mass-to-light ratio of 10, rotation velocity from NGC 7603 of 200 km/s 
in the outer disc, and a distance to the satellite of 40 kpc (equivalent
to an angular distance of 1 arcminute), we would need a galaxy
with B$\sim 17$. Where is this galaxy? 
Unless we suppose that NGC 7603B is this galaxy,
things are not easy to explain; and, of course, it would imply that we
have an anomalous redshift.

\begin{figure}
\includegraphics[width=6cm]{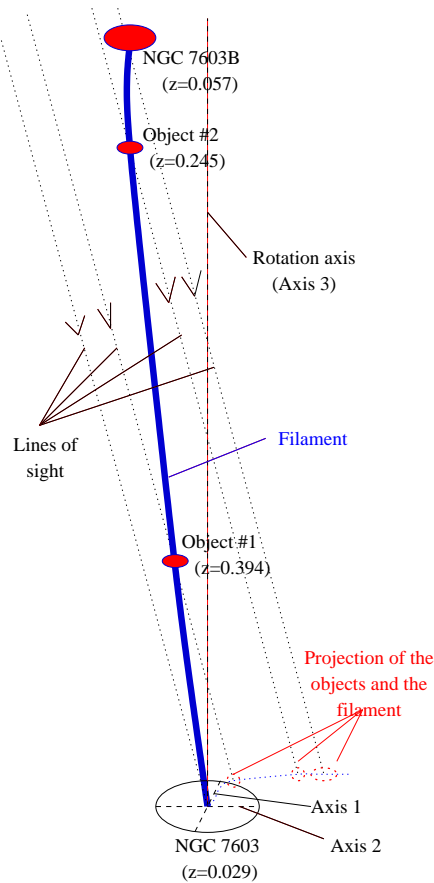} 
\caption{Possible representation of the system of NGC 7603 and other surrounding galaxies if we accept
the hypothesis that the three last objects were ejected by the parent galaxy NGC 7603. 
The inclination of NGC 7603 with respect the line of sight is $\approx 20^\circ $. 
The major axis  in the projected image (Axis 1) has a position
angle $\approx -15^\circ $; the minor axis in the projected image is 
``Axis 2''. Reproduction of Fig. 8 of \cite{Lop04}.}
\label{Fig:model}
\end{figure}

Therefore, some facts, although not conclusive, seem to suggest that there is an
interaction between the four galaxies of different redshift: the existence
of the filament itself, the strong H$_\alpha $ emission apparently observed
in the HII galaxies typical of dwarf galaxies, and the low probability
of having three background sources projected on to the filament
(see  below for a discussion of probabilities). 
As a speculative hypothesis, we might think that the three galaxies were
ejected by NGC 7603. There is no unique representation of the system in terms of 
this model of ejection. We do not have  enough information about the distances of the sources 
with respect to the parent galaxy to build a unique 3-D representation.
For instance, Fig. \ref{Fig:model} represents a possible configuration according to
the ejection theory. The inclination of the galaxy is around 20 deg with respect the 
line-of-sight (ellipticity $\approx 0.35$), so slight
deviations of the objects from the rotation axis could produce the projected
image that we have observed. Fig. \ref{Fig:model} represents a model in which
the filament is not in the plane of the galaxy, but is ejected in a direction nearly
perpendicular to the plane.

\section{NEQ3}

\begin{figure}[htb]
\includegraphics[width=7cm]{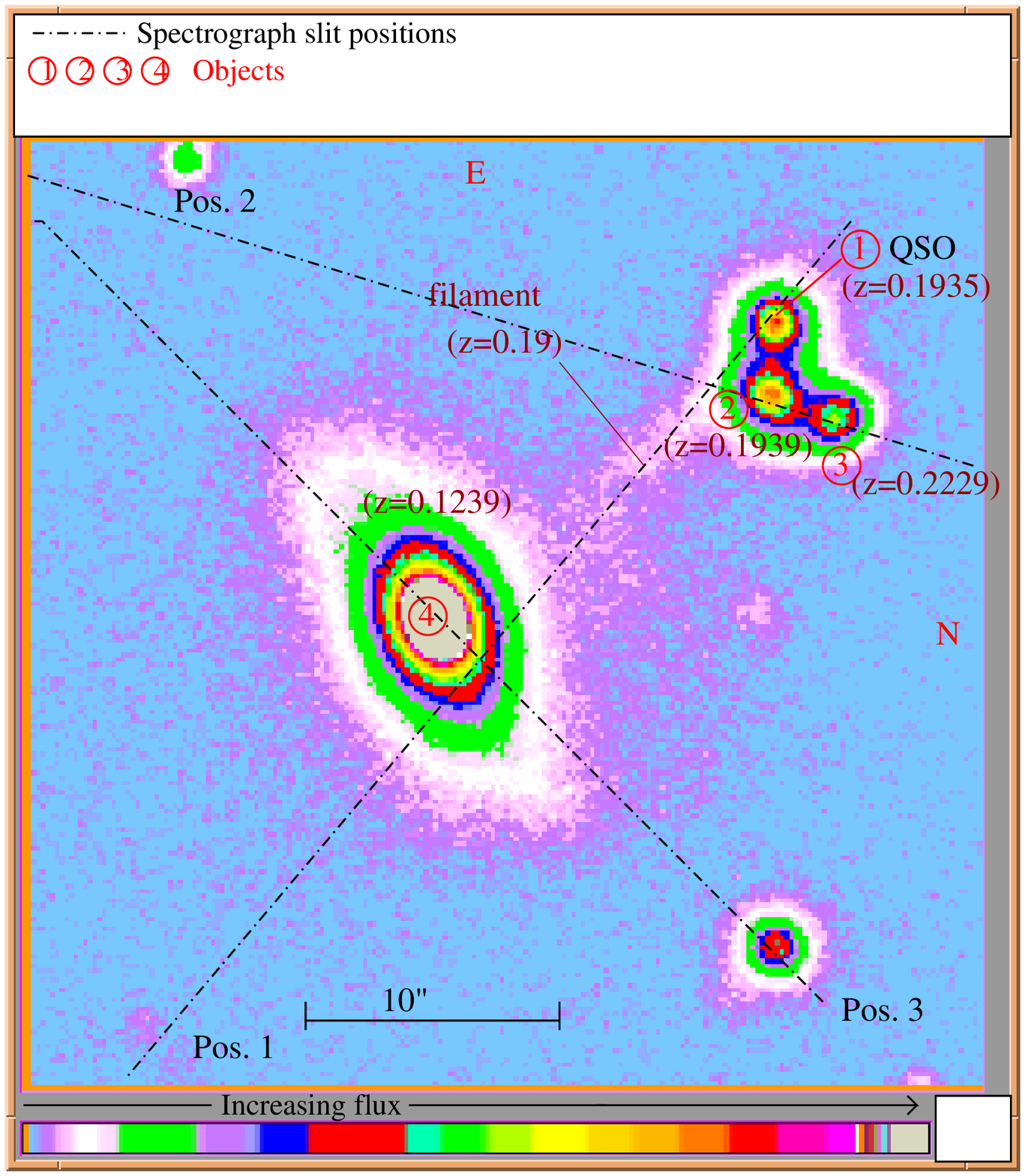} 
\caption{A false color-scale in the $r^\prime$-Sloan band of the region around the
system NEQ3 (only a section  $\sim 40^{\prime\prime}\times 40^{\prime\prime}$ has been
plotted). The plot also shows the position of the slits used in \cite{Gut04}.
Taken on the 2.5 m NOT (La Palma, Spain). Reproduction of Fig. 1 of
\cite{Gut04}.}
\label{ima1}
\end{figure}

The NEQ3 system  (see Fig. \ref{ima1})
comprises three compact-like objects with relative separations of 
2.6 and 2.8 arcsec with respect to the central object,  all of them lying along the 
minor axis of an apparent lenticular galaxy at $\sim 17$ arcsecs. A filament is
situated along the optical line connecting the main galaxy and the
three compact objects. Although  a rather intriguing system, surprisingly the
only study of it before our paper\cite{Gut04} was published 27 years ago
by Arp\cite{Arp77}, who measured the redshift of the main galaxy as
0.12, and two of the compact objects as 0.19; Arp claimed it to be an interesting
case of anomalous redshift. Apparently, little interest was generated concerning
 this problem since the system was forgotten for such a long time.

We have obtained \cite{Gut04}  a better image of the filament 
(previously noted by Arp) along the line of
the minor axis of object 4 (see Fig. \ref{ima1}). It also seems that the filament decreases
its brightness from object 1--2 to object 4. Object 4 looks fairly
symmetric with the morphology  typical of a lenticular (S0/Sa) galaxy,
and its spectrum presents only absorption lines characteristic of an old population. 
The only signs of distortion are the filament in the direction towards
objects 1--3, and an extended emission to the north which seems to
enclose a weak object ($r^\prime$-Sloan=21.8 mag) situated at $\sim 13$
arcsec. The magnitudes in the Sloan $r^\prime$ band for objects 1--3 are
19.8, 19.6, 20.2,  17.3 for object 4 (without extinction correction).
The spectra of objects 1, 2 and 3 are dominated by emission lines whose identification allows the
unambiguous determination of their redshifts\cite{Gut04}: the main features are the
lines of the hydrogen Balmer series, and the lines of OII and OIII.
Other minor features like  SII and and NII are also detected.  
The position of all these lines is consistent  to within the
uncertainty of our spectral resolution. The resulting redshifts are
$0.1935\pm 0.0002$, $0.1939\pm 0.0005$, $0.2229\pm 0.0002$ (a new 
discordant redshift galaxy), and
$0.1239\pm 0.0005$ for objects 1, 2, 3 and 4 respectively.
Object 1 has a typical broad line spectrum, while objects 2 and 3
have only narrow emission lines. According to their line ratios, we have classified 
object 1 as a QSO/Seyfert 1 galaxy, and objects 2 and 3 as HII 
galaxies.
The equivalent H$_\alpha $ equivalent widths of objects 2 and 3 are also relatively high,
66 and 52 \AA \ respectively, but not as much as in the knots of the filament of NGC 7603.  
The spectrum of the filament is very noisy and   has no  obvious
features. However, we tentatively identify a maximum in the cross-correlation 
function with a template spectrum that corresponds approximately to a redshift of 
0.19. A secondary maximum also appears  at $z=0.12$. 

Therefore, again, as in NGC 7603, we have seen that the system is 
even more anomalous than previously thought: 
 we now have three different redshifts instead of two.
Also as in NGC 7603, the origin of the filament is a mystery; it is supposed to
be due to the interaction of the pair 1,2 with some other galaxy to the south-west.
Where is this object? 
It seems that object 4 is the galaxy concerned, and this would imply 
anomalous redshift. It is also possible that the  interaction between 
objects 1 and 2 might produce the filament, and it would be a
coincidence that it pointed towards object 4.
The association of objects 1 and
2 is also particularly interesting since they constitute one of the few known  
galaxy/QSO pairs with a very small angular separation (2.6 arcseconds), 
the galaxy is undergoing an intense burst of star formation, and there
is a diffuse filament possibly associated with the pair. All of this evidence seems
to be an indication of a strong interaction between the QSO and the HII galaxy, objects with
more or less the same luminosity.
In a conventional scenario the role of objects 3 and 4 is unclear. For instance 
could the difference in redshift between objects 1 and 2, and object 3  ($\sim
0.03$) be produced by a difference in peculiar velocity? The difference would be
9,000 km s$^{-1}$. As far as we know, interactions between
galaxies with such a large difference have not been observed and 
would be difficult to explain within the framework of models of galaxy
formation. Finally, how to interpret in a conventional scenario the presence of several
asymmetries of the main galaxy, and a filament pointing outwards in 
the direction of the line connecting three compact objects with the 
centre of object 4?

\section{GC 0248+430 and B2 1637+29}

\begin{figure}[htb]
\includegraphics[width=8cm]{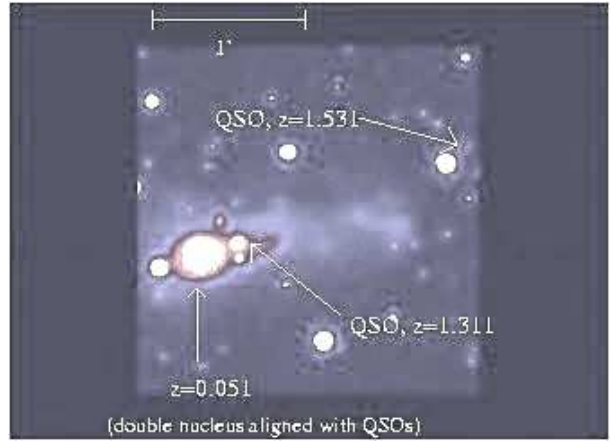} 
\caption{GC 0248+430, a galaxy with a double nucleus, and two QSOs in its
field. Sloan r' filter. Taken on the 2.5 m (La Palma, Spain).}
\label{Fig:GC0248}
\end{figure}

\begin{figure}[htb]
\includegraphics[width=12cm]{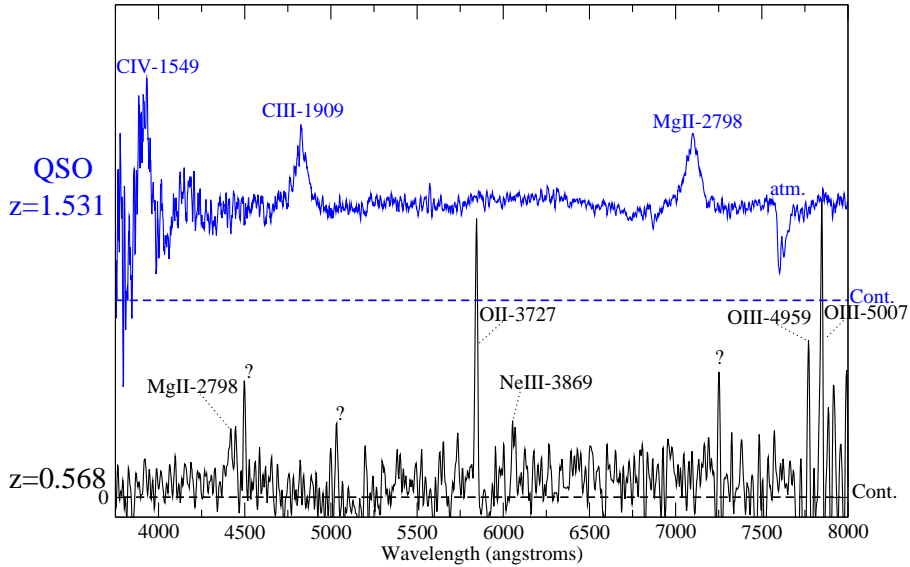} 
\caption{Spectra of the QSO with $z=1.531$ in the field of GC 0248+430
(taken with the 4.2 m WHT, La Palma, Spain)
and the emission-line object with $z=0.568$ in the field of B2 1637+29
(taken with 2.5 m NOT, La Palma, Spain.
Some lines marked  `?' are possibly contamination in the sky subtraction).}
\label{Fig:spectra}
\end{figure}

Fig. \ref{Fig:GC0248} shows the double nucleus 
galaxy GC 0248+430 and the field surrounding
it. The first quasar with $z=1.311$, $m_R=17.45$ was studied 
previously\cite{Wom90,Bor91,Hoy96} (our photometry gave 
$m_{r'}=18.80$ for observation on 2002 December 3; 
the difference is possibly due to variability), 
and the remarkable coincidence was
that the double nucleus of the galaxy and the QSO lie along a line.
Moreover, the gas extends in the direction of this QSO and beyond it,
and the galaxy is disturbed, due presumably to some interaction.
The new discovery to add to all this was the presence of a second QSO
with redshift $z=1.531$, $m_{u'}=21.55$, $m_{g'}=21.11$
$m_{r'}=20.76$ (see Fig. \ref{Fig:GC0248}).
The discovery was made by means of a systematic search for excess-UV sources
with  Sloan u'g'r' photometry on the NOT and spectroscopy with the WHT
to confirm the candidates
(method to detect QSOs explained in subsection 3.2 of \cite{Lop04}).
This object had been previously classified as a star from a photometric survey 
\cite{Kir94} but is undoubtedly   a QSO according to
the spectra with broad lines of CIV, CIII and MgII (see Fig. \ref{Fig:spectra}).
The position angle of the first QSO with respect to the major nucleus of the
galaxy is -73$^\circ $ at a distance of 14.4" from the major nucleus;
and the position angle of the second QSO is 
-68$^\circ $ at a distance of 108$''$ from the major nucleus. 
The position angle of the line which joins the two nuclei
of the galaxy is -77$\pm 18 ^\circ $ (the error bar is large because
both nuclei are very close---2.7$''$---and it was not possible to determine
 the position of the second centre accurately). 
Even if we forget the second nucleus and the gas ejection, all
in this direction\cite{Hoy96}, given the low density
of expected background quasars, the coincidence of the near-alignment
(the difference is 5 degrees) seems unlikely.

\begin{figure}[htb]
\includegraphics[width=8cm]{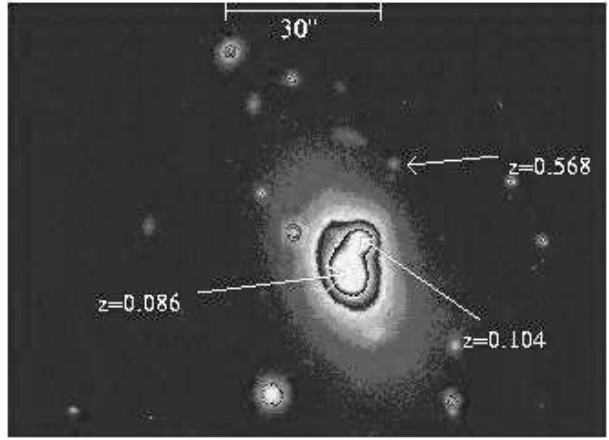} 
\caption{B2 1637+29, a system of extended objects separated by 4100 km/s, 
and a QSO in its field perfectly aligned with both (0.6 degrees of 
difference). Taken on the NOT.}
\label{Fig:B21637}
\end{figure}

Fig. \ref{Fig:B21637} shows the system B2 1637+29, either a galaxy
with two nuclei separated by 4100 km/s\cite{Rui98} 
or two separate galaxies, and the field surrounding it. 
We have serendipitously discovered  another galaxy with 
$z=0.568$, $m_{u'}=19.94$, $m_{g'}=20.75$, 
$m_{r'}=20.56$ (SDSS). We detected the emission lines of
OII, NeIII, OIII, and MgII (broad?, it is not clear; the classification
as QSO/Seyfert 1 is not discarded although, in principle, we simply claim that
it is an emission line galaxy; from the images it looks like an extended object); 
see Fig. \ref{Fig:spectra}. 
There is a fainter object which connects
the isophote of this object with those of B2 1637+39, but it was
too faint to derive its redshift.
The position angle of the line which joins the two nuclei
of the galaxy is -22.0$^\circ $ and the separation is 6.6$''$.
The position angle of the second galaxy with respect to the major nucleus of the
galaxy is -22.6$^\circ $ at a distance of 22.4$''$ from the major nucleus,
so there is  a difference of only 0.6 degrees (from SDSS astrometry, the number increases
to 1.1 degrees).
Indeed, the detection of this new object was totally fortuitous:
we placed the slit to take the spectra of both objects separated by
4100 km/s and by chance we took the spectrum of this object.

\section{VV172}

\begin{figure}[htb]
\includegraphics[width=8cm]{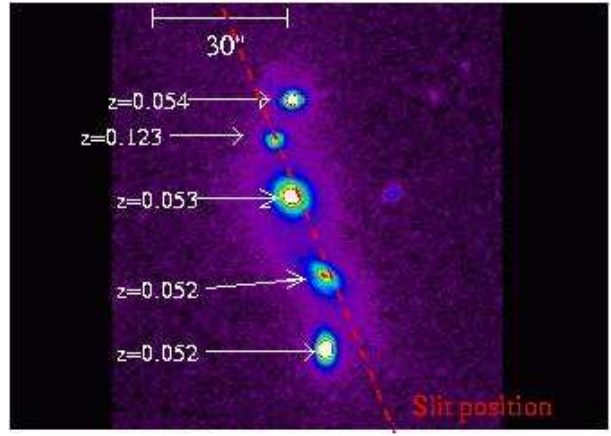} 
\caption{VV172, public-domain image in the R-band from the 3.6 m CFHT.}
\label{Fig:VV172}
\end{figure}

VV172 is a well known case of anomalous redshift in one of the members
of a Hickson Group. As shown in
Fig. \ref{Fig:VV172}, it consists of a chain of 5 galaxies in which
one has a discordant redshift (UGC 6514 E with 
$z=0.123$ instead of $z=0.052-0.054$ for the
other galaxies). The object with the discordant redshift is an
emission line galaxy, either an HII galaxy or a LINER, according
to our spectrum (why is it that
most of the galaxies in scenarios of discordant redshift are 
emission-line galaxies?); it is also very blue ($B-V=0.89$ \cite{Sul83},
plus 0.42 for the K-correction). It is either a blue compact or a spiral, 
perhaps a spiral because some structure was observed in their images. 
It has been analysed previously and was a matter
of discussion during the '70s and early '80s; later on it was apparently
forgotten. The extent and the smoothness of its enveloping halo, especially
in the vicinity of the discordant galaxy, was the strongest direct evidence
for interaction\cite{Sul83}. The
halo could not be explained by the simple overlapping of four or five normal
galaxy envelopes, suggesting that a significant part of the halo is
composed of stars governed by the overall gravitational potential of
the group\cite{Sul83}. 

\vspace{2cm}
\begin{figure}[htb]
\includegraphics[width=10cm]{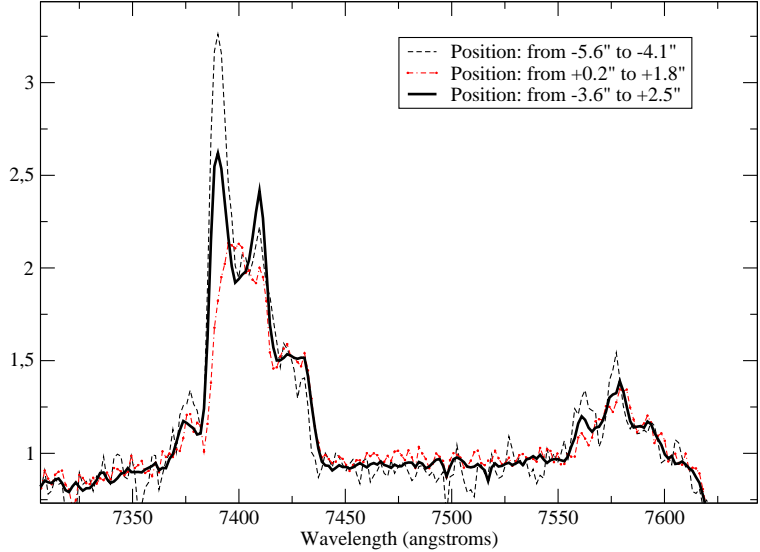} 
\caption{Spectrum of the object with $z=0.123$ in the  VV172 chain
(taken with the WHT).
Position is referred to the centre of this galaxy and  is positive
towards the south along the slit line, and negative towards the north along the
same line.}
\label{Fig:spectra2}
\end{figure}

Our research on this system was not  very intensive, 
and further observations
and analysis will be necessary. We took several spectra in the region,
among others the slit position marked in Fig. \ref{Fig:VV172}. The spectra
in the intergalaxy halo did not reveal  any redshift (there were no
emission lines, and the possible absorption lines were too faint to be detected).
However, we realized from the  spectrum of the discordant redshift galaxy
that it is a peculiar object. In Fig. \ref{Fig:spectra2} we see that the 
wavelengths corresponding to 
H$_\alpha $+NII-doublet have four lines (left, around 7400 \AA \ ), 
and the range of the SII-doublet (right, around 7575 \AA \ ) has three lines.
There are too many lines. We could also see H$_\beta $ asymmetrically broadened towards
redder wavelengths; OII and OIII did not present significant broadening.
All this information indicates that the lines
are duplicated (we do not see all of them because there is superposition
of the two sets of lines) and we have two components
with a difference of velocity of $\approx +700$ km/s in the same galaxy, possibly
a high velocity outflow: the main galaxy is ejecting some material, 
which is precisely what we are observing with an excess of $\approx 700$ km/s. 
This would indicate that ejection of matter is taking place in this galaxy.
A curiosity perhaps, but this adds a further anomaly to the previously
known one. Outflows are not very unusual, but outflows of
700 km/s are quite rare (only one galaxy of a sample of 75 
starburst-driven superwinds has a velocity higher than 700 Km/s\cite{Vei04}).

\section{MCG 7-25-46 and Stephan's Quintet}

\begin{figure}[htb]
\includegraphics[width=8cm]{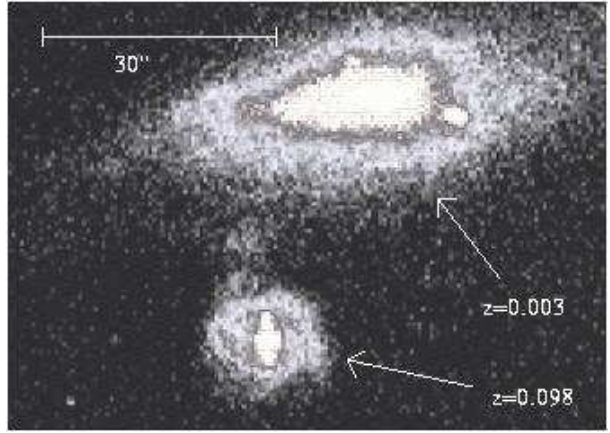} 
\caption{MCG 7-25-46 (or UGC 7175). Image in R-band taken with the 2.2 m
telescope at Calar Alto.}
\label{Fig:UGC7175}
\end{figure}

\begin{figure}[htb]
\includegraphics[width=8cm]{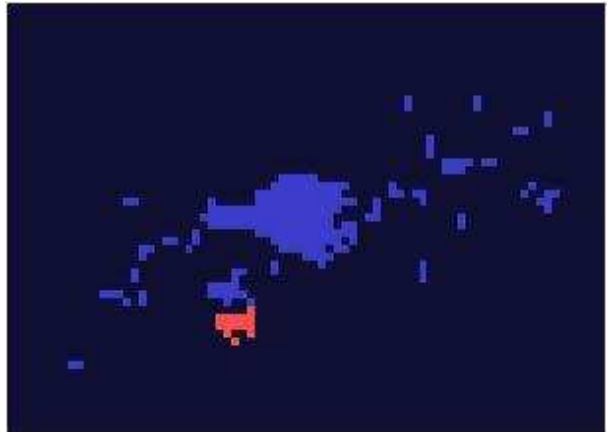} 
\caption{MCG 7-25-46 (or UGC 7175). Map of \protect{H$_\alpha $} velocities 
with two different velocities: $z=0.003$ (the main galaxy and the bridge),
$z=0.098$ (the small galaxy). H$_\alpha $ images obtained in
0.8 m IAC80 telescope (Tenerife, Spain) and 2. 2m Calar Alto telescope
(Spain).}
\label{Fig:UGC7175v}
\end{figure}

MCG 7-25-46 (or UGC 7175) (Fig. \ref{Fig:UGC7175}) was also analysed by Arp \cite{Arp77}:
a system with two galaxies connected by a bridge and with different
redshift: $z=0.003$ for the main galaxy and $z=0.098$ for the small one.
From our analysis, it is relevant that the bridge has the same redshift
as the main galaxy (see Fig. \ref{Fig:UGC7175v}). This had already been observed
by \cite{Sul85}, but they took a single slit spectrum, and we could
produce a 2D map of the H$_\alpha$ emission. \cite{Sul85} also observed
perturbation, due possibly to interaction, in the higher redshift
object. As in NGC 7603, one
could ask why MCG 7-25-46 ejects a filament/bridge in the direction
of the discordant redshift companion and not in other directions.
There is the further  interesting observation that the H$_\alpha $
emission at $z=0.003$ finishes exactly where the H$_\alpha $
emission of the galaxy with $z=0.098$ begins (which is supposed to
be in the background), there is no  overlap in the two emissions. 
Is it not a strange coincidence? 

This coincidence  reminds us of
the case in Stephan's Quintet, and the H$_\alpha $ bridge connecting
NGC 7320 to the other galaxies\cite{Gut02} that we have analysed with
the same technique: it also happens in this case that the H$_\alpha $ 
(Fig. 4 of \cite{Gut02}) with
discordant redshift begins exactly where the major component redshift
finishes in the bridge connecting  NGC 7320 to the rest of the group;
in such a case,  there was no overlapping of both H$_\alpha $ emissions
with different velocities. A coincidence? Perhaps. There are also other
coincidences in Stephan's Quintet, such as radio emission isophotes
with 6600--6700 km/s tracing quite exactly the shape of NGC 7320 
($\approx 800$ km/s) (Figs 5, 8 of \cite{Wil02}) and connecting it
with the rest of the Quintet, similar to the case of
NGC 3067+3C232 \cite{Car92}. None the less, there are other observations
 suggesting that there is no connection (see \cite{Gut02} and references
therein) and it is quite likely also that NGC 7320 is a foreground galaxy
with some coincidences in the system.

\section{Probabilities of being background/foreground galaxies}

There are two possible interpretations of these data: either the galaxies
with different redshift are objects with different distances and the
configurations are due to chance, or there are non-cosmological redshifts
and objects with different redshifts are at the same distance. The
first position, the standard one, defends the hypothesis that in all cases the main galaxy
is surrounded by background/foreground objects. The idea is quite straightforward.
The position of anomalous redshifts is not naive enough to deny this
possibility, and this might be the case in some examples. However, the
question is not whether such a fortuitous projection is ``possible''
but whether it is is ``probable''.

For the calculation of this probability, we assume that the background/foreground
objects are distributed according to a Poissonian distribution with the
average density in any line of sight.
There may be some clustering, but this does not essentially affect  the numbers.
A conspiracy in which our line
of sight crosses several clusters of galaxies at different redshifts is not justified 
because the increase in  probability due to the increase of  density
in lines of sight with clusters is compensated for by the additional factor to 
be multiplied to the present amount $P$ to take into account the probability 
of finding clusters in the line of sight. 
On average, in any  arbitrary line of sight in the sky,
the probability will be given anyway by the Poissonian calculation of $P$
(see further details in subsection 5.3.1 of \cite{Lop04}).

\subsection{Some clarifications concerning the typical
rebuff ``it is just an a posteriori calculation''}

Before showing some rough calculations on statistical probability, 
we would like to address the question of ``a posteriori probabilities''.
It is said that one should not carry out a calculation of the probability for 
an a priori known configuration of objects (for instance, that they form a 
certain geometrical figure) because, in some way, all possible configurations 
are peculiar and unique. We agree while we speak 
about random configurations which do not indicate anything special. 
For example, if the Orion constellation is observed and we want 
to calculate the chance of their stars being projected
in that exact configuration, we will get a null probability 
(tending towards zero as the allowed error in the position of the stars
with respect the given configuration goes to zero),
but the calculation of this probability is worthless because
we have selected a particular configuration observed a priori.
Therefore, the statistics to be carried out should not be about 
 the geometrical figure drawn by the sources, unless
that geometrical configuration is representative of a physical process
in an alternative theory (for instance, aligned sources might be 
representative of the ejection of sources by a parent source). 

In this last sense, we think that
much of the statistics already published is valid and indicates
the reality of some kind of statistical anomaly.
It would be useful to look out for 
physical representations indicating peculiarities beyond mere uniqueness.
We disagree with the claim that all attempts to calculate probabilities of
unexpected anomalies are a posteriori whose validity may therefore be rejected.
Some astrophysicists, when looking at our images, 
argue along the lines that  it is curious that some of our objects fall on the 
filament, but that since they do the probability is 1 and there is therefore nothing
special about our galaxies. According to this
argument, everything is possible in a Poissonian distribution and nothing
should surprise us. But we believe that statistics
is something more serious than the  postmodern rebuff that anything is possible. 

We think that this anti-statistical position, this way of rejecting the validity
of the calculated probabilities, is equivalent to the scepticism that
those unfamiliar with mathematics express when we discuss
the low probability of winning the lottery. They
continue to bet regardless with hope that, however low the probability,
somebody is sure to win so why not me.
Typically they are unaware of how low some probabilities are
and make no distinction between a case such as $P\sim 10^{-2}$, which is a low 
but certainly makes a win possible from time to time, and the
case $P\sim 10^{-7}$, which virtually ensures  no wins during seven lifetimes
of daily betting. Small numbers, like the huge numbers prevalent in astronomy,
are not easily assimilated. Of course, somebody wins the lottery
but this is because the number of players multiplied by the probability
to win each one is a number not much lower than one; otherwise, nobody
would ever be likely to win.

Even worst, imagine that a person wins the lottery four consecutive times
with only one bet each time.
If we did not believe in miracles, we might think that this person had cheated.
We might carry out some statistical calculations and show how improbable
it was that he/she could have won by chance. What might somebody say 
about these calculations,  
that they are not valid because they were carried out a posteriori (after
the person won the lottery four consecutive times)? We would not agree
because there is an alternative explanation (he/she is cheating; and this 
explanation could be thought of before the facts) and the event of winning
the lottery four consecutive times, apart from being unique among the random
possibilities, would be an indication to support this hypothesis.

For our cases, we will use facts (alignments, galaxies on to filaments) 
which suggest that an alternative theory
(a priori) claiming that galaxies/QSOs may be ejected by galaxies better
represents the observations. We will not fix the type of 
extragalactic object observed; neither will we calculate the probability
for a given type of galaxy, except when it is a QSO, since the alternative
hypothesis suggest that these objects have a close relationship with the
ejection scenario. 
We are not going to measure probabilities
of forming triangles or any shape observed a priori only because it
was observed. The peculiarity
that we want to analyse is not comparable with the previous example of
Orion because we have in mind a physical representation rather than
a given distribution of sources.
The difference with the Orion problem 
is that the peculiarity of Orion is not associated
with any peculiar physical representation to be explained by an alternative 
theory. The question is as follows: what is the probability, $P$, 
that the apparent fact be the fruit of a random projection of sources at
different distances? In other words, what is the probability, 
$P$, that the standard theory can explain the observed facts without aiming 
at alternative scenarios?

In the following subsections, we will take some information a priori
before the calculation, for instance the magnitude of the
object as a limiting magnitude of the probability density. We recognize
that it is in this sense an a posteriori calculation, we are calculating
the most pessimistic case (the lowest probability).
Some other authors (e.g. \cite{Slu03} hypotheses H2-H3) use in their 
calculation of the probabilities the limiting magnitude of the survey 
instead of the magnitude of the object, 
which gives a much higher probability. However, this is not totally correct 
because in case that the magnitude
of the object is much brighter than the limiting magnitude one should multiply
$P$ by a factor that characterizes the probability of this object being
much brighter than the limiting magnitude (the brighter it is, the lower
is the probability), and this is equivalent to using the magnitude of the
object. So we think that hypotheses H2-H3 in \cite{Slu03} 
are inappropriate because they lose statistical information 
(the existence of objects much brighter than the limiting
magnitude, which has  a low probability).

There are also other ways to lose information. One is to mix a datum 
with signal together with many other data without signal. This is what one gets 
when one uses a very large area instead of the area near the galaxies where the
excess of sources is found, for instance, or the area of the filament,
representative of the connection. Precisely because of that we
use the distance of the source to derive the limiting probability density, 
or the area of the filament where the galaxies are embedded 
when it is a relevant fact instead of the area of the whole image.
Using the radius of the image, for instance, would lose information on
the concentration of objects around the center or in the filament. 
If the distribution is homogeneous (as expected in background 
sources which have nothing to do with the parent galaxy) both radii should be 
more or less the same; a factor of two in average would differentiate them. 
If it turned out that the QSOs/galaxies are much more concentrated 
around the parent galaxy than at some distance from it, this would be meaningful 
and both ways of making the statistical calculations would be equivalent, but in
the first case one has to multiply the way of calculation 
by another probability which reflects how probable it is that 
in a large image we have QSO/galaxies near the 
galaxy instead of the borders of the image.

Taking into account all these considerations, we think that the following
values of $P$ might be slightly underestimated (by a factor not higher than 
10--100) with respect to an a priori calculation without any information on
magnitudes or radii, but values of $P$ lower than $\sim 10^{-4}$
should in any case be considered as statistically anomalous.
In order to do a fairer estimate of the probability, we could calculate
$P^*=2^n\times P$, where $n$ is the number of parameters on which $P$
depends. For instance, when we observe a source with magnitude 19
and we calculate $P(m<19)$ we are putting the limiting magnitude exactly
at the observed number; a fairer calculation would be $P^*(m<(19+x))$ such
that a source with magnitude is a typical average source in the range 
$m<(19+x)$, i.e. roughly that half of the sources with $m<(19+x)$ have 
$m<19$ and the other half have $19<m<(19+x)$. 
This is equivalent to calculating $P^*(m<(19+x))=2\times P(m<19)$
and we multiply by a factor two for any independent
parameter. Values of $P^*$ lower than $\sim 10^{-3}$ are to be
considered as statistically anomalous.

\subsection{NGC 7603}
\label{.probab}
 
From Figs \ref{Fig:f1} or \ref{hst}
it seems extremely unlikely that objects 1--4 at different distances
can, by chance, give a projection in the way these figures show up.
The probability is as follows: NGC 7603 has a filament of area $A$.
The probability of having 3 further independent sources with the
corresponding magnitudes of the objects 1--3 projected on that filament is
(assuming that the individual probabilities for each event, $p_i$, follow 
$e^{p_i}\approx 1$):

\begin{equation}
P_=\frac{A^3N_1(m\le m_1)N_2(m\le m_2)N_3(m\le m_3)}{3!}
\label{p1}
,\end{equation}
where $N_i$ is the source density on the sky for the type of sources
of the object $i$ with apparent magnitude less than $m_i$
(magnitudes corrected for Galactic+filament extinction, in order to be 
comparable with the galaxy counts in other fields), for the filter
in which we know the magnitude of the source. We will use, for instance,
filter B, but the statistics will give similar results for any filter.

The filament is approximately 35 arcsec long by 
4 arcsec wide (the area plotted in Fig. \ref{hst}):
$A\approx 35"\times 4"=140\ {\rm arcsec}^2$. Thus, with the magnitudes
and the observed complete (non-biassed)
galaxy counts\cite{Lop04}: $P\sim 3\times 10^{-9}$. 
$P$ depends on 5 parameters (3 limiting magnitudes and the width and height
of the area, so $P^*\sim 10^{-7}$, still low).
The probability would be lower if we consider that
 objects \#1 and \#2 are
positioned where the filament contacts NGC 7603B and NGC 7603 respectively,
or the distribution of redshifts (from major to minor), 
or we restricted the types of galaxies, especially the fact that
HII galaxies have a low probability of having a large 
H$_\alpha $ equivalent width \cite{Lop04}.
We do not consider these facts (also to avoid some claim of ``a posteriori'' 
calculation, see above) but even so the probability is low.
Even if we made a complete analysis of these characteristics among all
Seyfert galaxies like NGC 7603 (there are 237 AGN-galaxies in the whole sky 
with B magnitude less than 14.0, the magnitude of
NGC 7603\cite{Vau91}, according to SIMBAD), the probability
of having at least one case would be $\sim 7\times 10^{-7}$ 
[$P^*\sim 4\times 10^{-5}$], or an
order of magnitude larger if we consider all galaxies (Seyfert or otherwise)
up to magnitude 14.

\subsection{NEQ3 and MCG 7-25-46}

For NEQ3, we can make a single probabilistic
calculation starting from the pair formed by objects 1 and 2 in Fig. \ref{ima1}
and computing the probability of having object number 3 at 2.6 arcsec,
obtaining $P_1\sim 4\times 10^{-3}$. The probability of having the main galaxy ($R=17.3$
mag) at a distance of 17 arcsec along a filament (with a width of 2.5 arcsec) 
connecting objects 1-2 and 4 is $P_2\sim 5\times 10^{-4}$. 
Therefore, the probability of having all of them is
$P=P_1\times P_2/2!\sim 10^{-6}$ [$P^*\sim 3\times 10^{-5}$]. 
If we want to know how probable it is to find a system like
this in the whole sky we should multiply $P$ by the number of existing QSO-HII pairs  separated by
distances $\le 2.8''$. There is no precise information about this, but the number 
should not be very large since the number of cases like this that we know of is very 
low; in fact, it is quite rare.

The probability of MCG 7-25-46 having a galaxy with $m_B\sim 18$ 
(with a background density around 20 deg$^{-2}$\cite{Met91}) projected 
randomly onto its filament of area  $\sim 200$ arcsec$^2$ (10"x20") 
towards south is $P\sim 3\times 10^{-4}$ [$P^*\sim 2\times 10^{-3}$]. Since
the number of galaxies like MCG 7-25-46 ($m_B\approx 15$) is of the order of a
few ten of thousands in the whole sky, it is quite possible to find
a coincidence like this and it is statistically very likely that this is not an
anomalous redshift system. Nonetheless, as said, the origin of the filament
and the abrupt transition of H$_\alpha $ remains to be explained.

\subsection{GC 0248+430 and B2 1637+29}

The cases of possible ejection can be evaluated as the probability of having 
a source within a radius multiplied
by the probability of its being within a small range of position angles (except for 
the first pair of objects which defines the reference position angle).

For GC 0248+430, the probability of having the first QSO up to a 
b$_j$-magnitude $\sim 19$ (background density around 
2 deg$^{-2}$\cite{Boy00}) at distance up to 14.4$''$ is $P_1\sim 10^{-4}$.
The probability of having the second quasar up to a $b_j$-magnitude 
$\approx 21.3$ (background density around 50 deg$^{-2}$\cite{Boy00}) at a
distance of up to 108$''$ and within a position angle of range 20 degrees
(5 degrees per quadrant) in the whole 360 degrees is
$P_2\sim 0.14\times \frac{20}{360}=8\times 10^{-3}$ (not so low; it 
is possibly a background source). The global probability is
$P=P_1\times P_2/2!\sim 4\times 10^{-7}$ [$P^*\sim 10^{-5}$]. 

For B2 1637+29, the probability of the galaxy with a difference
of 4100 km/s from a background one, within 6.6$''$ up to a 
B-magnitude $\sim 18$ (background density around 20 deg$^{-2}$\cite{Met91}) 
is $P_1\sim 2\times 10^{-4}$. 
The probability of having the second up to  $B$-magnitude 
$\approx 21.0$ (density around 800 deg$^{-2}$\cite{Met91}) at a
distance of up to 22.4$''$ and within a position angle range of 2.4 degrees
(0.6 degrees per quadrant) in the whole circle is
$P_2\sim 0.1\times\frac{2.4}{360}=6\times 10^{-4}$.
The global probability is $P=P_1\times P_2/2!\sim 6\times 10^{-8}$ 
[$P^*\sim 2\times 10^{-6}$].
If we took the SDSS number of 1.1 degrees of difference in the position
angle, $P\sim 10^{-7}$ [$P^*\sim 3\times 10^{-6}$].

Again, we should multiply these
probabilities by the number of sources like GC 0248+430 (double nucleus 
sources up to the magnitude of this source), or B2 1637+29 in order to know how probable
is finding a configuration like this in the whole sky. 
Let us forget the fact of the double nucleus and consider the faintest
object (B2 1637+29, $m_B=16.5$). There are around $2\times 10^{5}$ galaxies
up to this magnitude\cite{Met91} in the whole sky. 
Only the fact that we have 2 cases like these with probabilities
$4\times 10^{-7}$ and $6\times 10^{-8}$-$10^{-7}$ among the 200 thousand galaxies
has a probability around $5\times 10^{-4}-10^{-3}$ ($P^*$ of the order
of unity). The probability is not low, so a 
background projection of some of the objects is quite
probable, but if many other systems with alignment are found, as said 
in the first section, the
probability will go lower and lower.

\subsection{VV172 and Stephan's Quintet}

The fact of observing a Seyfert galaxy with a relatively high velocity
outflow is rare in itself, but we will not consider it for the statistics in this study, 
only the fact of having a galaxy within a radius 10$''$ of the north-east galaxy
of the chain up to magnitude
$m_B=18.0$ (SIMBAD; density of galaxies of any type up to this magnitude
is around 30 deg$^{-2}$\cite{Met91}) gives
$P=7\times 10^{-4}$ [$P^*=3\times 10^{-3}$]. 
The same calculation for NGC 7320 in Stephan's Quintet
to have this galaxy with $m_B=13.8$ (SIMBAD; density of galaxies of any type up to this magnitude
is around 0.2 deg$^{-2}$\cite{Met91}) within a radius of 100$''$ from NGC 7319
gives $P=5\times 10^{-4}$ [$P^*=2\times 10^{-3}$].

There are 100 Hickson Groups,
and the probability of having at least one case like VV 172 among them is 
$P=0.07$ [$P^*=0.3$] or $P=0.05$ [$P^*=0.2$] for Stephan's Quintet, 
which is high enough to be a chance projection.
However, the problem is that there are 43 cases among the Hickson
Groups with members of discordant redshifts \cite{Sul97}, and although VV172 
and many other cases (like Stephan's Quintet) 
may be random projection of foreground/background sources,
the global statistical analysis \cite{Sul97} remains to be solved in 
standard terms. If the probability were the same for all systems ($P^*\sim
10^{-3}$), the probability of having 43 coincidences among 100 cases\footnote{The
number of cases should be indeed larger, because groups of 3 galaxies 
should be considered in the statistics rather than $\ge 4$ galaxies,
because they can give a group of 4 objects when a discordant object is
joint. In any case, the number would be a few hundred within the limiting
magnitudes of Hickson's criteria.} would be extremely low. 
Moreover, some other coincidences in VV172 and
Stephan's Quintet indicated in the corresponding sections above  were not
taken into account in this calculation, so the oddity of these systems
is still considerable. The case of an outflow with a probability $\sim 1/75$
among starbursts in VV172 deserves further attention.

\section{Gravitational lensing}

An explanation for anomalous redshift systems 
might be found in principle if we considered some kind of 
gravitational lensing by the foreground object. However,
the effect in the enhancement of the probability 
produced by an individual galaxy should be small.
Some rough calculations can illustrate this argument:
given a galaxy, the enhancement
in the density of background objects as a function of angular distance,
$\theta $, to the mass centre will be \cite{Wu96}:

\begin{equation}
q_Q(\theta )=\frac{N[m<m_{b,lim}+2.5\log \mu (\theta )]}{N(m<m_{b,lim})}\frac{1}{\mu (\theta)}
\label{gravlens}
,\end{equation}
where $\mu $ is the magnification factor.
In order to increase at least an order of magnitude in $P$ per object, 
we would need an average enhancement of $\sim 10$ in density for 
each of the galaxies. With the counts from \cite{Met91}, 
this requires an average magnification of
$\mu (\theta )$ of $\sim 2\times 10^4$. 
It is clear from eq. (\ref{gravlens}) that
the density of sources does not increase so quickly, unless the counts 
increase extremely rapidly with the limiting magnitude, which is not our case. 
This is so because the enhancement in the source counts increases due to the 
flux increase of each source but decreases due to the area distortion, which 
reduces the number counts by losing the sources within a given area \cite{Wu96}.
A magnification of $\sim 2\times 10^4$ is extremely high and
impossible to achieve by a galaxy lens. The highest known values of $\mu$ are 
up to a factor $\sim 30$ \cite{Ell01} for background objects apparently close to 
the central parts of massive clusters. Moreover, a single galaxy would only
produce a significant magnification at very close distances from the centre -
a few arcseconds.

The possibility that multiple minilenses are distributed in the halo of
the galaxy has also been proposed: gravitational mesolensing by King objects\cite{Bar97,Buk03}. 
Strong gravitational lensing would be produced by King lenses: globular clusters\cite{Buk03}, 
dwarf galaxies, or clusters of hidden mass with masses between 10$^3$ and 10$^9$ M$_\odot $.
This is an interesting idea, although we are not convinced by the proof presented by one of authors
of the idea \cite{Buk01} revealing excesses of galaxy/QSO pairs  with $z_{gal}>0.9z_{QSO}$,
because many of these pairs were indeed the same object classified both as QSOs and galaxies.
Anyway, the idea is interesting, and it might be considered as a serious proposal
to solve the statistical correlations between QSOs and galaxies in large surveys,
although, for individual cases,
the area where the magnification is important is very small, so again the probability
of having a large number of sources is most probably small. Further research is necessary in this
direction, but up to now no solution has been found in terms of gravitational lensing applied
to individual systems.

Weak gravitational lensing by dark matter has also been proposed as the cause of the
statistical correlations between low and high redshift objects, but 
this seems to be insufficient to explain them \cite{Kov89,Zhu97,Bur97,Bur01,Ben01,Gaz03,Jai03}, 
and cannot work at all for the correlations 
with the brightest and nearest galaxies. 
More recently, Scranton et al.\cite{Scr05} have contradicted 
these results and have claimed that the correlation between QSOs and galaxies 
from the SDSS-survey is due to weak gravitational lensing. Indeed, what they 
have found was an ad hoc fit of the halo distribution function to an
angular cross-correlation with very small amplitude ($\omega _{GQ}<0.04$) 
of faint galaxies with QSO candidates
selected photometrically (5\% of this sample are not QSOs; \cite{Ric04}). 
Who knows what  the origin is of this very small\footnote{Small 
because the mean separation among galaxies is small
and any positive correlation of QSOs around a galaxy would be diluted with
the contamination of many other QSOs possibly belonging to other galaxies.} 
cross-correlation?
In any case, as said, no explanation of gravitational lensing for
correlations with the brightest and nearest galaxies is possible in terms 
of gravitational lensing, for instance for the high amplitude
angular correlation found by Chu et al. \cite{Chu84}: $\omega _{GQ}\sim 5$.
Scranton et al., even if they were right, have not solved the question of the 
correlation of galaxies and QSOs, because cross-correlations with bright 
and nearby galaxies, which are the most significant, are still without 
explanation in standard cosmological terms.

\section{Other problems with QSOs redshifts}

Other anomalies or questions with not very clear answers have arisen about
the nature of QSOs, along with the suspicion that their distance is not as high as 
indicated  by the cosmological interpretation of their redshift.
The luminosity required for QSOs to be at such large distances is
between $10^{43}$ to $10^{47}$ erg/s, an enormous energy to be produced
in a relatively compact region (to justify the strong variability in short 
times). Although this problem is solved by means
of megahuge black holes, the explanation might also be related to a bad
determination of the distance.
It seems that there are special problems in justifying the abundance of very 
high luminosity QSOs at $z\sim 6$, and the gravitational lensing solution 
does not work\cite{Yam03,Fan01b}. Moreover, the huge dispersion in the 
magnitude--redshift relation for QSOs observed
from the Hewitt \& Burbidge catalog\cite{Hew87} makes it impossible to derive a
Hubble law for them. This is a not strong argument since the intrinsic 
dispersion of luminosities might be high itself, but 
it might possibly indicate that something is wrong with 
the distance measurement. 

Another caveat is that
QSOs with very different redshifts have a low 
apparent brightness dispersion. This must be explained in the standard scenario
as QSOs evolving their 
intrinsic properties so that they get smaller and fainter as the universe 
evolves. QSOs with the highest redshift reach equivalent absolute magnitudes 
in optical filters (with K-correction) of -30 or brighter, while
local QSOs have absolute magnitudes around -16 (e.g. SDSS J113323.97+550415.8
with a redshift of only $z=0.0084$). What a strange way to evolve for QSOs,
that they should become so faint exactly in our epoch!

Superluminal motions of distant sources ($D$) 
are observed, i.e. angular speeds $\omega $ between 
two radio-emitting blobs which imply linear velocities $v=D\omega $ greater
than the speed  of light\cite{Coh86}. There are explanations for this.
The so called relativistic beaming model\cite{Ree67} assumes that there is one blob,
$A$, which is fixed while blob $B$ is traveling almost directly towards
the observer with speed $V<c$ with an angle $\cos ^{-1}(V/c)$ between
the line of advance and the line $B$-observer. This leads to an apparent
velocity of separation which may be greater than $c$. There is also another proposal 
in a gravitational bending scenario\cite{Chi79}. However,
both explanations share the common criticism of being contrived and
having a somewhat low probability ($\sim 10^{-4}$)\cite{Nar84}.

If the cosmological distance of a QSO is correct, then its huge radiation
would be strong enough to ionize the intergalactic medium. However, there
are QSOs like PG 0052+251 at the core 
of a normal spiral galaxy in which this host galaxy appears undisturbed 
by the QSO radiation\cite{Bah96}.

Other problems are that the mean Faraday rotation 
is less for objects at $z\approx 2$ than at $z\approx 1$ 
instead of increasing as expected
if the polarization of radio emission rotates when it passes through magnetized 
extragalactic plasmas; the metallicity and dust content of very high redshift 
QSOs and their host galaxies is equal or even larger in some cases than 
their metallicity at low redshifts, etc. (see \cite{Lop03}, subsect. 2.3.2).
In general, it seems that a non-cosmological redshift solution for QSOs fits 
many observations better .

\section{Discussion}

Some of the cases described in this paper may be just fortuitous 
cases in which background objects are close to the main galaxy, 
although the statistical mean correlations remain to be explained, and some 
lone objects have a very small probability of being a projection of background 
objects. Gravitational lensing seems not to be a solution yet, although
further research is required, and the aim that
{\it the probabilities be calculated a posteriori} is not in general 
an appropriate answer to avoid/forget the problem.

There are two possibilities: either we have been extremely lucky and all
cases with $P^*\sim 10^{-3}-10^{-8}$ were simply fortunate coincidences,
or there are at least some few cases of non-cosmological redshifts. 
If we accept that some of our objects (maybe not all of them) 
with different redshifts had the distance of the main galaxy, there might
be some truth in  those models \cite{Bur99b,Arp99a,Arp99b,Arp01,Bel02a,Bel02b} 
in which QSOs and other types of galaxies are ejected by a parent
galaxy. In these models, galaxies beget galaxies, not all the galaxies
would be not made from 
initial density fluctuations in a Big Bang Universe.
The narrow line character in some of these objects
would be a result of the ejection and interaction with the filament.
Evidence is shown in other papers \cite{Kee98,Kee99,Arp99a,Bur03} 
that when QSOs interact with 
ambient material they become less
compact and have narrower lines emitted from a more diffuse body. 
This could be the physical explanation.
Dynamically-disturbed starburst galaxies, as illustrated by
the case of NGC 2777 \cite{Arp88}, tend to be the small companions of larger 
nearby galaxies belonging to older stellar populations. According to
Arp\cite{Arp88}, they are 
recently created galaxies in which star formation is stimulated by recent
ejection from the parent galaxy; it has been suggested that some older stars, together with 
stellar material, are removed from the larger galaxy in the course of this ejection. 
Therefore, the observed fact of observing narrow emission line galaxies instead of QSOs is also
contemplated in the model, although the analysis of QSO--galaxy associations
is more frequent. The origin of these sources, through the interaction,
would also explain the high observed equivalent width in their H$_\alpha $ 
lines. 

For the explanation of the intrinsic redshift, there are several hypotheses. This
is not the place to discuss these, since this is a paper about the observational
facts (the phenomenology). Some proposals have been given at this meeting
CCC-I and there are others in the literature\cite{Nar89,Lop04}.

Summing up, observations challenge the standard model, which assumes that the 
redshift of all galaxies is due to the expansion of the Universe, and we must 
consider they are at least an open problem to be solved.


\begin{theacknowledgments}
Thanks are given to F. Prada (IAA, Granada, Spain) for providing us
an image of MCG 7-25-46 in H$_\alpha $ and R from the 2.2 m 
Calar-Alto telescope (Spain), to Rub\'en J. D\'\i az (C\'ordoba,
Argentina) for helpful discussions on outflows in relation with
our case in VV172, and to Terry Mahoney (IAC, Tenerife, Spain)
for proof-reading of this paper.

\end{theacknowledgments}





\bibliography{author}

\end{document}